\documentclass[twocolumn]{aastex7}
\usepackage{graphicx}   
\usepackage{subfigure}
\usepackage{multirow}
\usepackage{amsmath}
\usepackage{lineno}

\graphicspath{{./}{figures/}}

\shorttitle{baryon in cosmic filaments}
\shortauthors{Mo, Zhu, Yang, Zheng \& Feng.}
\begin{document}
% \linenumbers
\title{Constraining the Baryon Content of Cosmic Filaments Using Localized Fast Radio Bursts and DESI Imaging Data}

\author{Jian-Feng Mo}
\email{mojf5@mail.sysu.edu.cn}
\affil{School of Physics and Astronomy, Sun Yat-Sen University, Zhuhai campus, No. 2, Daxue Road \\
Zhuhai, Guangdong, 519082, China}
\affil{CSST Science Center for the Guangdong-Hong Kong-Macau Greater Bay Area, Daxue Road 2, 519082, Zhuhai, China}

\author[0000-0002-1189-2855]{Weishan Zhu}
\email{zhuwshan5@mail.sysu.edu.cn}
\affil{School of Physics and Astronomy, Sun Yat-Sen University, Zhuhai campus, No. 2, Daxue Road \\
Zhuhai, Guangdong, 519082, China}
\affil{CSST Science Center for the Guangdong-Hong Kong-Macau Greater Bay Area, Daxue Road 2, 519082, Zhuhai, China}

\author{Qi-Rui Yang}
\email{yangqr8@mail2.sysu.edu.cn}
\affil{School of Physics and Astronomy, Sun Yat-Sen University, Zhuhai campus, No. 2, Daxue Road \\
Zhuhai, Guangdong, 519082, China}
\affil{CSST Science Center for the Guangdong-Hong Kong-Macau Greater Bay Area, Daxue Road 2, 519082, Zhuhai, China}

\author{Yi Zheng}
\email{zhengyi27@mail.sysu.edu.cn}
\affil{School of Physics and Astronomy, Sun Yat-Sen University, Zhuhai campus, No. 2, Daxue Road \\
Zhuhai, Guangdong, 519082, China}
\affil{CSST Science Center for the Guangdong-Hong Kong-Macau Greater Bay Area, Daxue Road 2, 519082, Zhuhai, China}

\author{Long-Long Feng}
\email{flonglong@mail.sysu.edu.cn}
\affil{School of Physics and Astronomy, Sun Yat-Sen University, Zhuhai campus, No. 2, Daxue Road \\
Zhuhai, Guangdong, 519082, China}
\affil{CSST Science Center for the Guangdong-Hong Kong-Macau Greater Bay Area, Daxue Road 2, 519082, Zhuhai, China}

\correspondingauthor{Weishan Zhu}
\email{zhuwshan5@mail.sysu.edu.cn}

%%\collaboration{all}{The Terra Mater collaboration}

\begin{abstract}
Cosmic filaments are thought to host a substantial fraction of the missing baryons at redshifts $z<2$. In this study, we constraint the baryonic content of these filaments using localized Fast Radio Bursts (FRBs). Filaments are identified from the galaxy distribution in the Dark Energy Spectroscopic Instrument (DESI) imaging surveys using the DisPerSE algorithm. We find tentative evidence ($\sim 3 \sigma$ significance) for a divergence in the relationship between the dispersion measure (DM) contributed by the intergalactic medium and redshift for FRBs whose signals intersect cosmic filaments compared to those that do not, suggesting excess baryons in the filamentary structures. Assuming an isothermal $\beta$-model gas profile with $\beta=2/3$, this discrepancy is best explained by a central baryon overdensity of $\delta_0 = 21^{+13}_{-12}$, broadly consistent with previous simulation and observational results. The inferred baryon fraction residing in filaments decreases with redshift, from approximately $0.25$--$0.30\,\Omega_b$ at $z=0.02$ to $0.15$--$0.30\,\Omega_b$ at $z=0.5$, and $0.03$--$0.04\,\Omega_b$ at $z=0.8$. These estimates are likely lower bounds, particularly at $z>0.5$, due to the limited number of identified filaments and localized FRBs at higher redshifts. We also examine various factors that may affect the statistical significance of our results. Our method offers an independent approach to tracing baryons in cosmic filaments and underscores the importance of expanding localized FRB samples and deepening galaxy surveys—key steps toward refining these estimates and addressing the missing baryon problem.

\end{abstract}



\section{Introduction} \label{sec:intro}

According to the $\Lambda \rm{CDM}$ cosmology (e.g., \citealt{2016A&A...594A..13P}), $\sim 5 \%$ of the energy density in the universe are made up by baryonic matter. However, a significant portion, about $30$--$50\%$, at redshift $z<2$ are `missing' from detection for a long time (\citealt{1992MNRAS.258P..14P};\ \citealt{1998ApJ...503..518F};\ \citealt{2012ApJ...759...23S}; \citealt{2016ApJ...817..111D}). Meanwhile, simulations predicted that approximately $40$--$60\%$ of the baryons reside within the filamentary structures of cosmic web ( \citealt{1999ApJ...514....1C};\ \citealt{2001ApJ...552..473D};\ \citealt{2016MNRAS.457.3024H}; \ \citealt{2017ApJ...838...21Z}; \ \citealt{2018MNRAS.473...68C}) at $z<2$. Furthermore, simulations indicate that cosmic filaments host approximately $60$--$90\%$ of the warm-hot intergalactic medium (WHIM), which is believed to account for the majority of `missing' baryons (\citealt{2019MNRAS.486.3766M,2021A&A...646A.156T}). 

A variety of observational techniques have been developed to detect baryons in cosmic filaments, including X-ray emission and absorption (e.g., \citealt{2009ApJ...699.1765B}; \ \citealt{2015Natur.528..105E}; \citealt{2017A&A...606A...1A}; \citealt{2002ApJ...572L.127F};\ \citealt{2005ApJ...629..700N};\ \citealt{2016MNRAS.457.4236B};\ \citealt{2018Natur.558..406N};\ \citealt{2019A&A...621A..88N}; \ \citealt{2020A&A...643L...2T};\citealt{2025A&A...698A.270M}), the thermal Sunyaev-Zel'dovich (SZ) effect (\citealt{2018A&A...609A..49B};\ \citealt{2019MNRAS.483..223T};\ \citealt{2019A&A...624A..48D}; \ \citealt{2020A&A...637A..41T}). 
However, the statistical significance of many of these detections remains at the $3$--$4\,\sigma$ level. In addition, several reported detections require further confirmation. Continued development of complementary observational tools is crucial for robustly identifying and quantifying the baryon content in cosmic filaments.

Fast radio bursts (FRBs) are luminous and millisecond-duration events, though their physical origin is still an open question (\citealt{2007Sci...318..777L,2019ARA&A..57..417C, 2023RvMP...95c5005Z}). As FRB signals propagate through ionized plasma, lower frequencies are delayed relative to higher ones—a phenomenon quantified by the dispersion measure (DM), which reflects the integrated free electron density along the line of sight (L.O.S.). \cite{2014ApJ...780L..33M} first proposed using the DM of FRBs to probe the universe’s missing baryons, and subsequent studies have made significant progress.

\cite{2020Natur.581..391M} analyzed the DM of seven well-localized, low-redshift FRBs and measured the cosmic baryon density as $\Omega_b h_{70}=0.051^{+0.021}_{-0.025}$, broadly consistent with estimates from cosmic microwave background (CMB) observations \citep{2016A&A...594A..13P}, albeit with large statistical uncertainties. \citealt{2022ApJ...940L..29Y} used 22 localized FRBs to measure the baryon content in the universe, giving $\Omega_b=0.049^{+0.0036}_{-0.0033}$ in 1$\sigma$ confidence level. More recently, \cite{2025NatAs...9.1226C} further used a larger sample of 69 localized FRBs to yield $\Omega_b h_{70}=0.051^{+0.006}_{-0.006}$. Despite these advancements, such studies do not resolve the detailed distribution of baryons within collapsed halos and the cosmic web.

Based on cosmological hydrodynamical simulations, \cite{2018ApJ...865..147Z} demonstrated that ionized baryons in cosmic filaments contribute approximately $35$--$45\%$ of the total DM of FRBs at $z < 2$. Furthermore, \cite{2021ApJ...906...95Z} found that the median DM contribution from foreground halos is about $30\%$ of that from the intergalactic medium (IGM), albeit with substantially variance.
More recently, several studies have reported cases where FRB signals intersect individual foreground halos, groups, walls, or filaments (e.g., \citealt{2020ApJ...901..134S, 2023ApJ...954L...7L, 2023ApJ...954...71S, 2024ApJ...973..151K, 2024arXiv240514182F, 2024arXiv241007307S}). These developments make it increasingly feasible to estimate the specific DM contributions from ionized baryons in collapsed halos and cosmic web, offering a path toward constraining their detailed baryon distributions. 

With the growing number of well-localized events (e.g., \citealt{2017ApJ...834L...7T, 2020ApJ...903..152H, 2021ApJS..257...59C, 2023arXiv230703344L, 2025ApJS..277...43M}), FRBs are becoming a powerful tool to probe baryons in cosmic filaments. At $z<1$, filaments occupy $\sim10\%$ of the cosmic volume, and high-redshift FRBs can intersect multiple filaments. Due to the inhomogeneous filament distribution, even FRBs at similar redshifts may traverse different numbers of filaments, leading to variation in their extragalactic DMs. However, this signal is complicated by contributions from host galaxies, their CGM, and foreground halos (\citealt{2021ApJ...906...95Z, 2025ApJS..277...43M}). A statistically large sample of localized FRBs is thus crucial to improve signal-to-noise when comparing DMs of FRBs that intersect filaments to those that do not.

This paper probes the baryonic content of cosmic filaments using localized FRBs. Section 2 outlines the data and methodology used to detect baryons in filaments and estimate their baryon content; Section 3 presents the results; Section 5 summarizes our conclusions. Factors affecting the statistical significance are discussed in the Appendix. We adopt cosmological parameters: $\Omega_m = 0.3089$, $\Omega_b = 0.0486$, $\Omega_\Lambda = 0.6911$ and $h = 0.6774$ \citep[][]{2016A&A...594A..13P}. 

\section{Methodology} \label{sec:method}

\subsection{Galaxy catalog from DESI Legacy Imaging Surveys and filaments }
\begin{figure}
    \centering
%\begin{centering}
    \hspace{-0.5cm}
    \includegraphics[width=0.7\columnwidth]{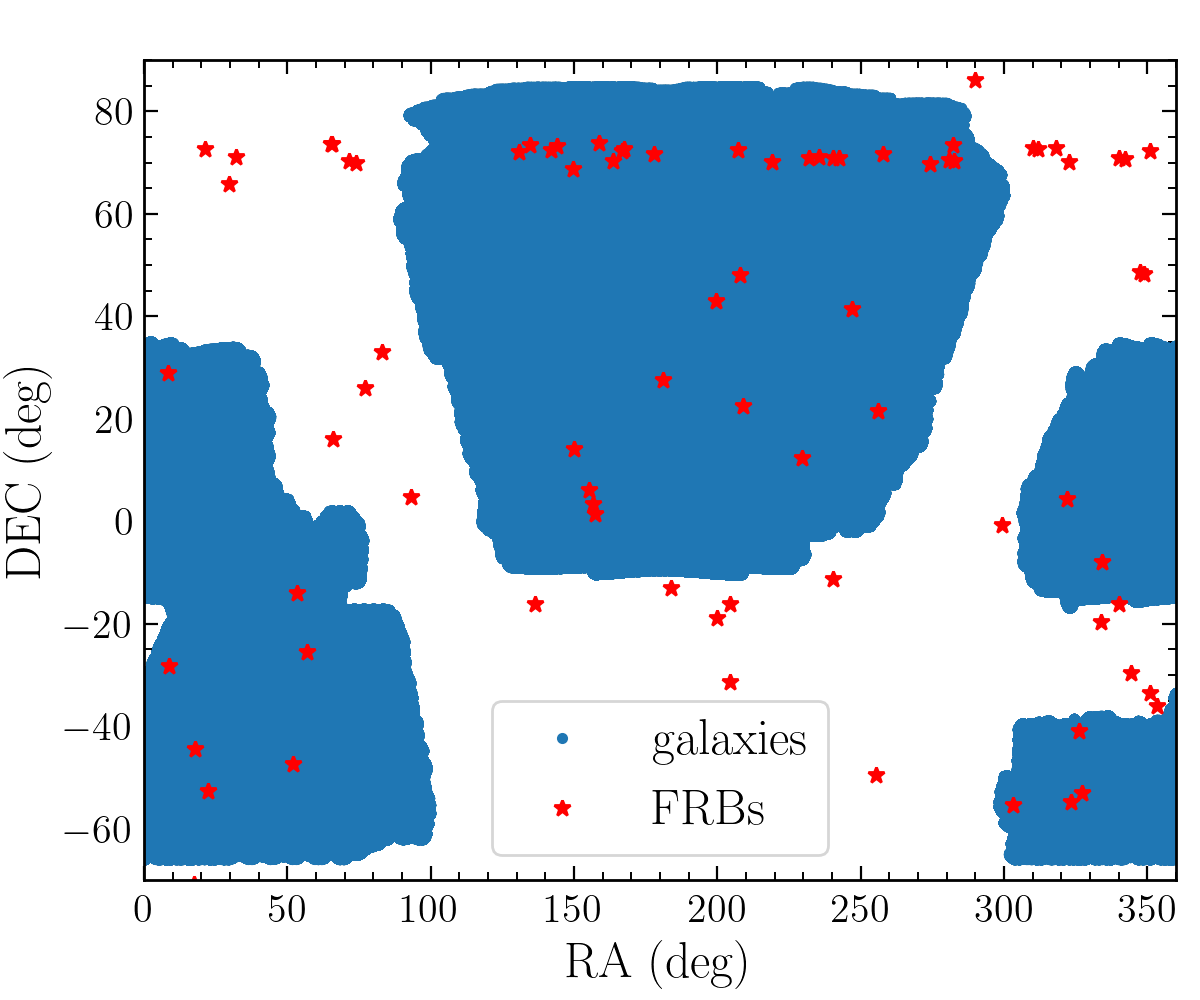}
    \includegraphics[width=0.8\columnwidth]{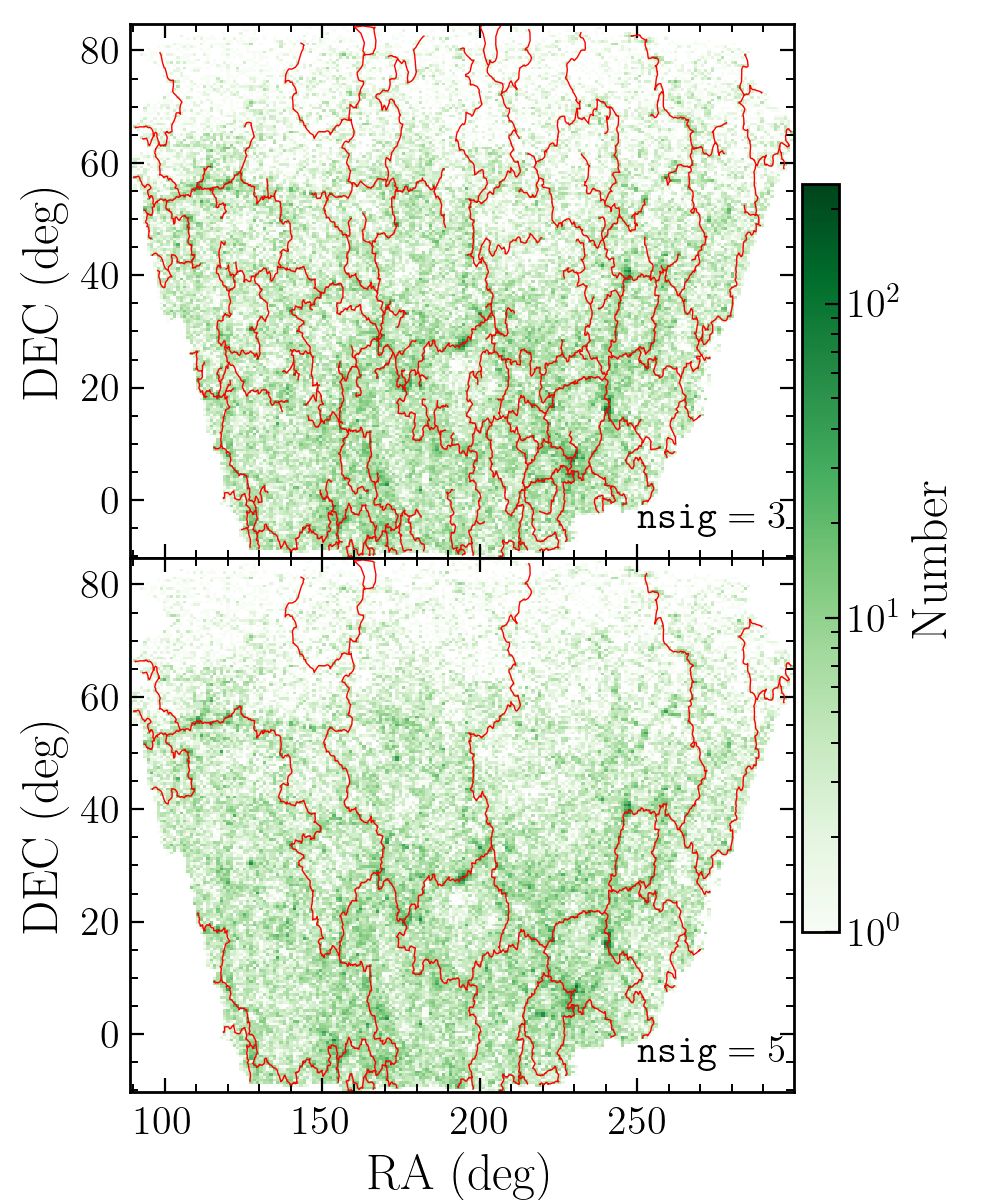}
    \includegraphics[width=0.8\columnwidth]{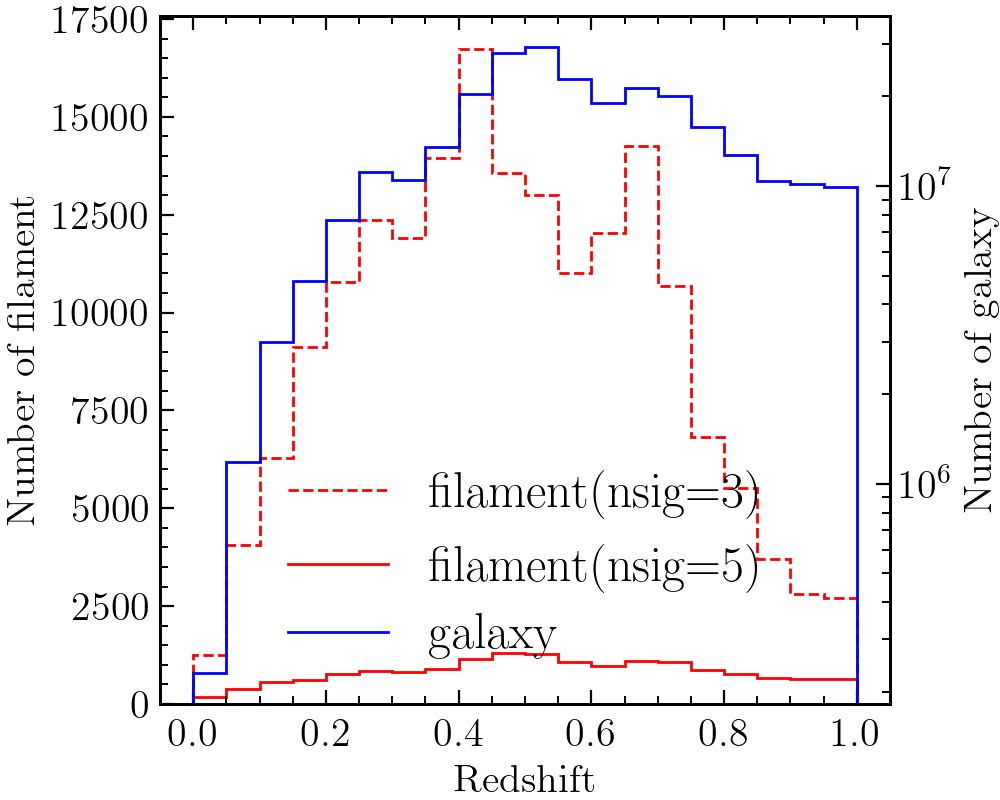}
    \caption{Top: The sky area (blue region) in the Dark Energy Spectroscopic Instrument (DESI) Legacy Imaging Surveys and positions of localized FRBs (red pentagrams). Middle: Filaments and galaxies at $z = 0$–0.05 across the northern sky. Red lines show filament skeletons identified by DisPerSE with \texttt{nsig}=3 and \texttt{nsig}=5 thresholds; green shading indicates galaxy density. Bottom: red lines show filament counts for \texttt{nsig}=3 (dashed) and \texttt{nsig}=5 (solid); blue line shows galaxy counts per redshift bin.}
%\end{centering}
    \label{fig:exampleFila}
\end{figure}

\cite{2019ApJS..242....8Z} published a catalog of photometric redshifts and stellar masses for $\sim$300 million galaxies from the DESI Legacy Imaging Surveys \citep{2019AJ....157..168D}, covering over 14,000 $\mathrm{deg^2}$. The galaxies span redshifts up to $z \sim 1$ and stellar masses from $10^{8.4}$ to $10^{11.9}\,\mathrm{M}_\odot$. The sky coverage is shown in blue in the top panel of Figure \ref{fig:exampleFila}.

%\begin{figure}
%    \centering
%    \includegraphics[width=1.\columnwidth]{RaDec_zou2019_FRB.png}
%    \caption{The galaxy distribution (blue region) in the Dark Energy Spectroscopic Instrument (DESI) Legacy Imaging Surveys and localized FRBs (red pentagrams) on the sky.}
%    \label{fig:radec_zou2019_frb}
%\end{figure}

Using this galaxy catalog, we identify cosmic filaments using the DisPerSE algorithm\footnote{\href{https://www2.iap.fr/users/sousbie/web/html/indexd41d.html?}{https://www2.iap.fr/users/sousbie/web/html/indexd41d.html?}} \citep{2011MNRAS.414..350S, 2011MNRAS.414..384S}, which is widely applied to both observations (e.g., SDSS; \citealt{2020A&A...642A..19M}) and simulations (e.g., MillenniumTNG; \citealt{2024A&A...684A..63G}) in both two-dimensional \citep[e.g.,][]{2024MNRAS.534.1682O} and three-dimensional \citep[e.g.,][]{2025arXiv250206484B, 2025ApJ...989..187Y} analyses.

Given the photometric redshift uncertainty of $\sim$0.02 in the DESI Legacy Imaging Surveys catalog \citep{2019ApJS..242....8Z}, we divide the galaxy sample into 20 redshift bins of $\Delta z = 0.05$ across $0 < z < 1$. In each bin, cosmic filaments are identified in 2D using DisPerSE, based on the galaxies’  right ascension (RA) and declination (DEC). DisPerSE constructs the density field with the Delaunay Tessellation Field Estimator (DTFE; \citealt{2000A&A...363L..29S, 2009LNP...665..291V}) and extracts filamentary structures via discrete Morse theory by identifying critical points (minima, saddles, and maxima) in the gradient field.

In DisPerSE, filaments are defined as pairs of field lines connecting saddle points to maxima. To filter out noise, DisPerSE applies persistence theory, which measures the topological significance of structures. Following \cite{2020A&A...642A..19M}, we adopt persistence thresholds of \texttt{nsig}=3 and 5, corresponding to 3$\sigma$ and 5$\sigma$ significance levels. While \texttt{nsig}=5 yields more robust filaments, it may exclude weaker but real structures. We use \texttt{nsig}=5 as the default and compare with \texttt{nsig}=3 to assess threshold sensitivity. Further implementation details are provided in the Appendix \ref{sec:appendix_class}.

The middle panel of Figure~\ref{fig:exampleFila} shows filament identification in the redshift range $0$–$0.05$ across the northern sky, using persistence thresholds of \texttt{nsig} = 3 (top) and \texttt{nsig} = 5 (middle), overlaid on the galaxy distribution. The bottom panel displays the number of galaxies and filaments per redshift bin. Filament identification becomes incomplete beyond $z > 0.5$ due to the decreasing number of galaxies in the DESI imaging survey at higher redshifts.

\subsection{Two groups of localized FRBs: intersected filaments or not} \label{subsec:filament classification}

We use 84 localized FRBs, combining the 71 events compiled by \cite{2025ApJS..277...43M} with 13 new ones—11 from the CRAFT survey \citep{2025PASA...42...36S} and 2 from CHIME/FRB \citep{2025ApJ...980L..24C, 2025ApJ...979L..21S}. Their sky positions are shown as red pentagrams in the top panel of Figure~\ref{fig:exampleFila}. Of these, 46 lie within the DESI area. From this subset, we select 37 high-confidence FRBs (with $\mathrm{P_{cc}} < 0.05$ or $\mathrm{P_{host}} > 0.95$) to test whether their lines of sight intersect cosmic filaments. The 37 FRBs are divided into two groups: `Pass', whose lines of sight intersect filaments from the galaxy catalog of DESI imaging surveys within redshift bins $j \cdot 0.05 < z < (j+1) \cdot 0.05$ ($j=0,\dots,19$), and `NoPass', which do not intersect any filament. The filament redshift in each bin is approximated as $z_{\mathrm{fila}}(j) \approx j \cdot 0.05 + 0.025$.

\begin{figure}
    \centering
    \includegraphics[width=0.8\columnwidth]{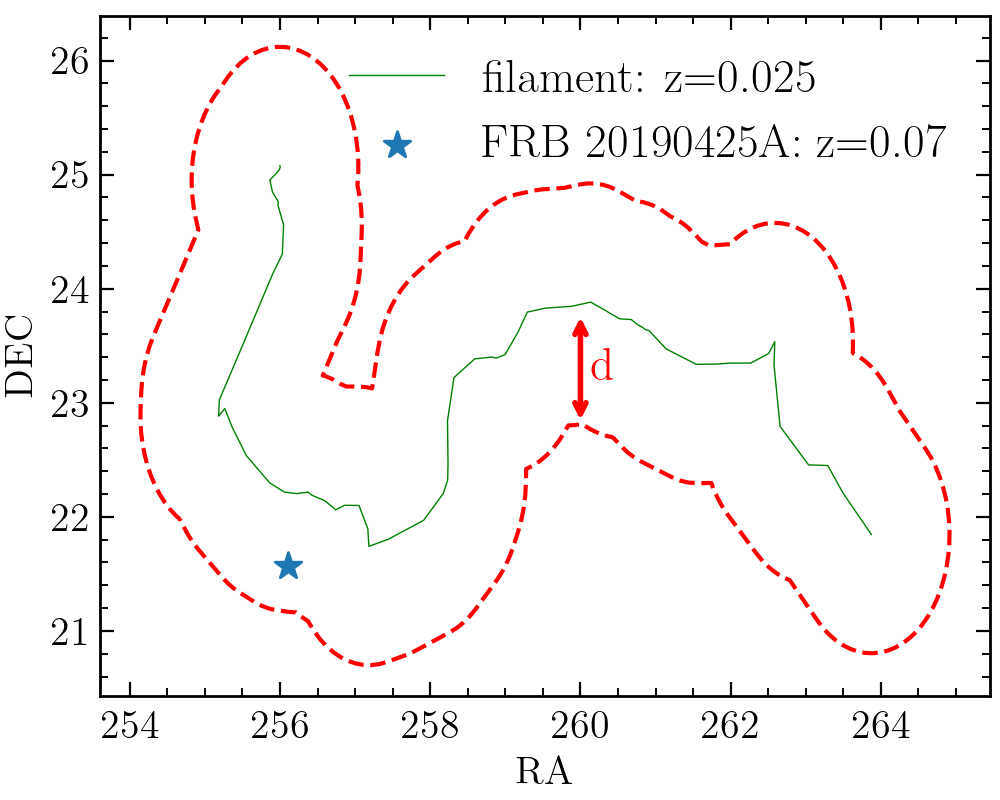}
    \includegraphics[width=0.8\columnwidth]{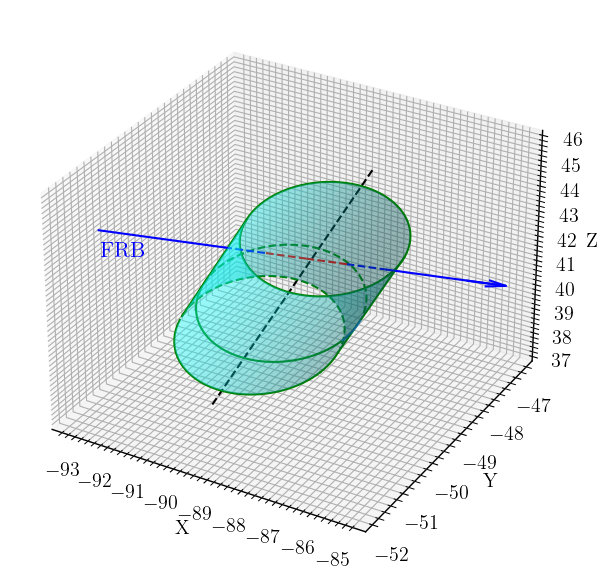}
    \caption{Upper: Example of an FRB intersecting a filament in 2D. The solid green line marks the filament spine, and `d' indicates its angular width. Bottom: 3D illustration of an FRB-filament intersection. The black dashed line shows the filament's skeleton, the cyan cylinder marks its region with radius $\mathrm{R_{fila}}$, the blue line is the sightlines to FRB, and the red segment indicates the intersecting path.}
    \label{fig:through fila 2d}
\end{figure}

To identify filaments intersected by FRB sightlines, we estimate their angular width as $d = \mathrm{R_{ini}}(1+z)/r_{\mathrm{com}}$, where $\mathrm{R_{ini}} = 2\,\mathrm{Mpc}$ is the initial estimate of filament radius and $r_{\mathrm{com}}$ is the comoving distance to the filament’s redshift bin. The choice of $\mathrm{R_{ini}} = 2\,\mathrm{Mpc}$ is motivated by \cite{2025ApJ...989..187Y}, who find most filaments are less than 2Mpc across. A filament is considered intersected if the FRB’s sky position falls within its projected angular width $d$ and its redshift exceeds that of the filament. This selection method is illustrated in the upper panel of Figure~\ref{fig:through fila 2d}.

To validate candidate filaments, we estimate their physical radius, $\mathrm{R_{fila}}$, using the correlation with stellar mass per unit length from \cite{2025ApJ...989..187Y}, then check if the FRB line of sight intersects the filament’s cylindrical region of radius $\mathrm{R_{fila}}$ in 3D space (see lower panel of Figure~\ref{fig:through fila 2d}). More details are provided in the Appendix \ref{sec:appendix_intersect}. FRBs meeting this criterion form the `Pass' subgroup. Among 37 securely localized FRBs, we find 7 intersect filaments in our default catalog, totaling 12 crossings. Table~\ref{tab:FRBs} lists their classifications, number of intersections, and filament redshifts. We also determine the specific intersection paths (e.g., red segment in the bottom panel of Figure~\ref{fig:through fila 2d}) used to model $\mathrm{DM_{fila}}$, the dispersion measure from baryonic gas in filaments, as described below.

\begin{deluxetable*}{cccccccc|ccc}
\tablecaption{The basic information of 37 securelylocalized FRBs and derived results of FRB-intersected filaments.}
\label{tab:FRBs}
\tablehead{
\colhead{} & \multicolumn{7}{c}{ Basic information of FRBs } & \multicolumn{3}{c}{Derived results of intersected filaments}  \\
\colhead{} & \colhead{FRB} & \colhead{RA} & \colhead{DEC} & \colhead{redshift} & \colhead{$\mathrm{DM_{obs}}$} & \colhead{$\mathrm{P_{cc}/P_{host}}$} & \colhead{Ref.} & \colhead{Number} & \colhead{$z_{\mathrm{fila}}$}  & \colhead{$\mathrm{DM_{fila}}$} \\ 
\colhead{} & \colhead{} & \colhead{degree} & \colhead{degree} & \colhead{} & \colhead{$\mathrm{pc\,cm^{-3}}$} & \colhead{} & \colhead{} & \colhead{} & \colhead{}  & \colhead{$\mathrm{pc\,cm^{-3}}$}
}
\startdata
\multirow{7}{*}{ `Pass' } & 190425A & 256.11 & 21.57 & 0.0715 & 127.8 & 0.0012/- & \cite{2023arXiv231010018B} & 1 & 0.025 & 27   \\
  &  200906A & 53.50 & $-$14.08 & 0.3688 & 577.8 & -/1 & \cite{2022AJ....163...69B} & 2 & 0.175,0.225 & 73   \\ %31,42
  &  220105A & 208.80 & 22.47 & 0.2784 & 583.0 & -/1 & \cite{2023ApJ...954...80G} & 2 & 0.025,0.075 & 56   \\ %29,27
  &  220509G & 282.67 & 70.24 & 0.0894 & 269.5 & -/0.99 & \cite{2023arXiv230703344L} & 1 & 0.025 & 32   \\
  &  221219A & 257.63 & 71.63 & 0.553 & 706.7 & -/0.99 & \cite{2024Natur.635...61S} & 3 & 0.025,0.125,0.175 & 103  \\ %25,38,40
  &  230307A & 177.78 & 71.70 & 0.2706 & 608.9 & -/0.97 & \cite{2024Natur.635...61S} & 2 & 0.175,0.225 & 78   \\ %35,43
  &  240114A & 321.92 & 4.33 & 0.1306 & 527.1 & -/0.997 & \cite{2025ApJ...980L..24C} & 1 & 0.075 & 35   \\
  \hline
\multirow{30}{*}{ `NoPass' } & 180924B & 326.10 & $-$40.90 & 0.3214 & 361.4 & 0.0018/0.9994 & \cite{2020ApJ...903..152H} &  \multicolumn{3}{c}{\multirow{30}{*}{-}}    \\
 & 181030A & 158.58 & 73.75 & 0.0039 & 103.5 & 0.0025/- & \cite{2021ApJ...919L..24B} &  &  &    \\
 & 181112A & 327.35 & $-$52.97 & 0.4755 & 589.3 & 0.0257/0.9274 & \cite{2019Sci...366..231P} &  &  &    \\ %\cite{2019Sci...366..231P} for pcc/phost?
 & 181223C & 181.08 & 27.58 & 0.0302 & 111.6 & 0.04/- & \cite{2023arXiv231010018B} &  &  &    \\
 & 190608B & 334.02 & $-$7.90 & 0.1178 & 338.7 & 0.0016/1 & \cite{2020ApJ...895L..37B} &  &  &    \\ % 2020ApJ...895L..37B for pcc/phost?
 & 191001A & 323.35 & $-$54.75 & 0.234 & 506.9 & 0.0031/0.9995 & \cite{2020ApJ...903..152H} &  &  &    \\
 & 200120E & 149.49 & 68.83 & $-$0.0001 & 87.8 & 0.0006/- & \cite{2021ApJ...910L..18B} &  &  &    \\
 & 200223B & 8.27 & 28.83 & 0.06024 & 203.8 & 0.01/0.899 & \cite{2023arXiv230402638I} &  &  &    \\
 & 200430A & 229.71 & 12.38 & 0.16 & 380.1 & 0.0051/1.0 & \cite{2020ApJ...903..152H} &  &  &    \\
 & 210117A & 339.98 & $-$16.15 & 0.214 & 730.0 & -/0.9984 & \cite{2023ApJ...948...67B} &  &  &    \\
 & 211212A & 157.35 & 1.36 & 0.0707 & 206.0 & -/0.998 & \cite{2022MNRAS.516.4862J} &  &  &    \\
 & 220204A & 274.22 & 69.72 & 0.4012 & 612.584 & -/0.99 & \cite{2024Natur.635...61S} &  &  &    \\
 & 220310F & 134.72 & 73.49 & 0.47796 & 462.24 & -/0.99 & \cite{2023arXiv230703344L} &  &  &    \\
 & 220418A & 219.10 & 70.10 & 0.622 & 623.25 & -/0.97 & \cite{2023arXiv230703344L} &  &  &    \\
 & 220914A & 282.06 & 73.34 & 0.1139 & 631.28 & -/0.97 & \cite{2023arXiv230703344L} &  &  &    \\
 & 220920A & 240.26 & 70.92 & 0.15824 & 314.99 & -/0.98 & \cite{2023arXiv230703344L} &  &  &    \\
 & 221012A & 280.80 & 70.52 & 0.28467 & 441.08 & -/1.0 & \cite{2023arXiv230703344L} &  &  &    \\
 & 221106A & 56.70 & $-$25.57 & 0.2044 & 343.8 & -/0.9708 & \cite{2025PASA...42...36S} &  &  &    \\
 & 230124 & 231.92 & 70.97 & 0.0939 & 590.574 & -/0.99 & \cite{2024Natur.635...61S} &  &  &    \\
 & 230526A & 22.23 & $-$52.72 & 0.157 & 361.4 & -/0.997 & \cite{2025PASA...42...36S} &  &  &    \\
 & 230626A & 235.63 & 71.13 & 0.327 & 452.723 & -/0.99 & \cite{2024Natur.635...61S} &  &  &    \\
 & 230628A & 166.79 & 72.28 & 0.127 & 344.952 & -/0.95 & \cite{2024Natur.635...61S} &  &  &    \\
 & 230708A & 303.12 & $-$55.36 & 0.105 & 411.51 & -/1.0 & \cite{2025PASA...42...36S} &  &  &    \\
 & 230712A & 167.36 & 72.56 & 0.4525 & 587.567 & -/0.99 & \cite{2024Natur.635...61S} &  &  &    \\
 & 230902A & 52.14 & $-$47.33 & 0.3619 & 440.1 & -/1.0 & \cite{2025PASA...42...36S} &  &  &    \\
 & 231226A & 155.36 & 6.11 & 0.1569 & 329.9 & -/1.0 & \cite{2025PASA...42...36S} &  &  &    \\
 & 240201A & 149.91 & 14.09 & 0.042729 & 374.5 & -/1.0 & \cite{2025PASA...42...36S} &  &  &    \\
 & 240210A & 8.78 & $-$28.27 & 0.023686 & 283.73 & -/1.0 & \cite{2025PASA...42...36S} &  &  &    \\
 & 240310A & 17.62 & $-$44.44 & 0.127 & 601.8 & -/0.9884 & \cite{2025PASA...42...36S} &  &  &    \\
 \enddata
\end{deluxetable*}

\subsection{Modeling of DM caused by baryons in intersecting filaments} \label{subsec:DMfila}

The DM in the rest frame is defined as the integral of the free electron number density $n_e$ along the L.O.S., i.e., 
\begin{equation}
% \begin{split}
    \mathrm{DM}=\int n_e dl.
    \label{eqn:DM definition}
% \end{split}
\end{equation}

For those FRBs whose sightlines intersect cosmic filaments, we estimate the DM caused by filaments using the gas density profile modeled as an isothermal single $\beta$ model with $\beta = 2/3$, following \cite{2021ApJ...920....2Z}, i.e., 
\begin{equation}
% \begin{split}
    \rho_{\mathrm{gas}}(r) = \frac{\rho_{\mathrm{gas},0}}{(1+(\frac{r}{r_c})^2)^{\frac{3}{2}\beta}} = \frac{\rho_{\mathrm{gas},0}}{1+(\frac{r}{r_c})^2} ,
    \label{eqn:gas density profile}
% \end{split}
\end{equation}
where the core radius is set to $r_c = 0.64\,\mathrm{R_{fila}}$, and the central gas density is $\rho_{\mathrm{gas},0} = (1 + \delta_0)\Omega_b \rho_{\mathrm{crit}}(1 + z)^3$, where $\delta_0$ is the central overdensity, $\Omega_b$ is the baryon density at $z = 0$, and $\rho_{\mathrm{crit}} = 3H_0^2 / (8\pi G)$ is the critical density; $H_0$ is the Hubble constant at $z=0$, $G$ is the Newtonian gravitational constant. According to \cite{2025ApJ...989..187Y}, we assume $\delta_0$ is redshift-independent.

Considering a hydrogen (H) mass fraction of $Y_{\mathrm{H}}=\frac{3}{4}$ and a helium (He) mass fraction of $Y_{\mathrm{He}}=\frac{1}{4}$, along with ionization fractions $\chi_{\mathrm{H}}\sim\chi_{\mathrm{He}}\sim1$ for filaments at redshift $z<3$ \citep{2014ApJ...783L..35D}, the resulting free electron number density in the filament is given by:
\begin{equation}
\begin{split}
    n_e(r) &= n_{\mathrm{H}} \cdot Y_{\mathrm{H}}\cdot \chi_{\mathrm{H}} + 2n_{\mathrm{He}}\cdot Y_{\mathrm{He}}\cdot \chi_{\mathrm{He}} \\
    &= \frac{\rho_{\mathrm{gas}}(r)}{m_p} \cdot Y_{\mathrm{H}}\cdot \chi_{\mathrm{H}} + 2\frac{\rho_{\mathrm{gas}}(r)}{4m_p}\cdot Y_{\mathrm{He}}\cdot \chi_{\mathrm{He}} \\ 
    &= \frac{\rho_{\mathrm{gas}}(r)}{m_p}(\frac{3}{4}+\frac{1}{2}\times \frac{1}{4})=\frac{7}{8}\frac{\rho_{\mathrm{gas}}(r)}{m_p},
    \label{eqn:ne}
\end{split}
\end{equation}
where $m_p$ is the mass of proton. 

The DM contributed by an intersecting filament, $\mathrm{DM_{fila}}$, is computed using Equation~\ref{eqn:DM definition}, with the integration path defined as in the previous section (see red line in the bottom panel of Figure~\ref{fig:through fila 2d}).

\subsection{Estimation of baryon mass in filaments}
\label{subsec:estimate baryon}

\begin{figure*}
\hspace{0.5cm}
\begin{centering}
    \includegraphics[width=0.9\textwidth]{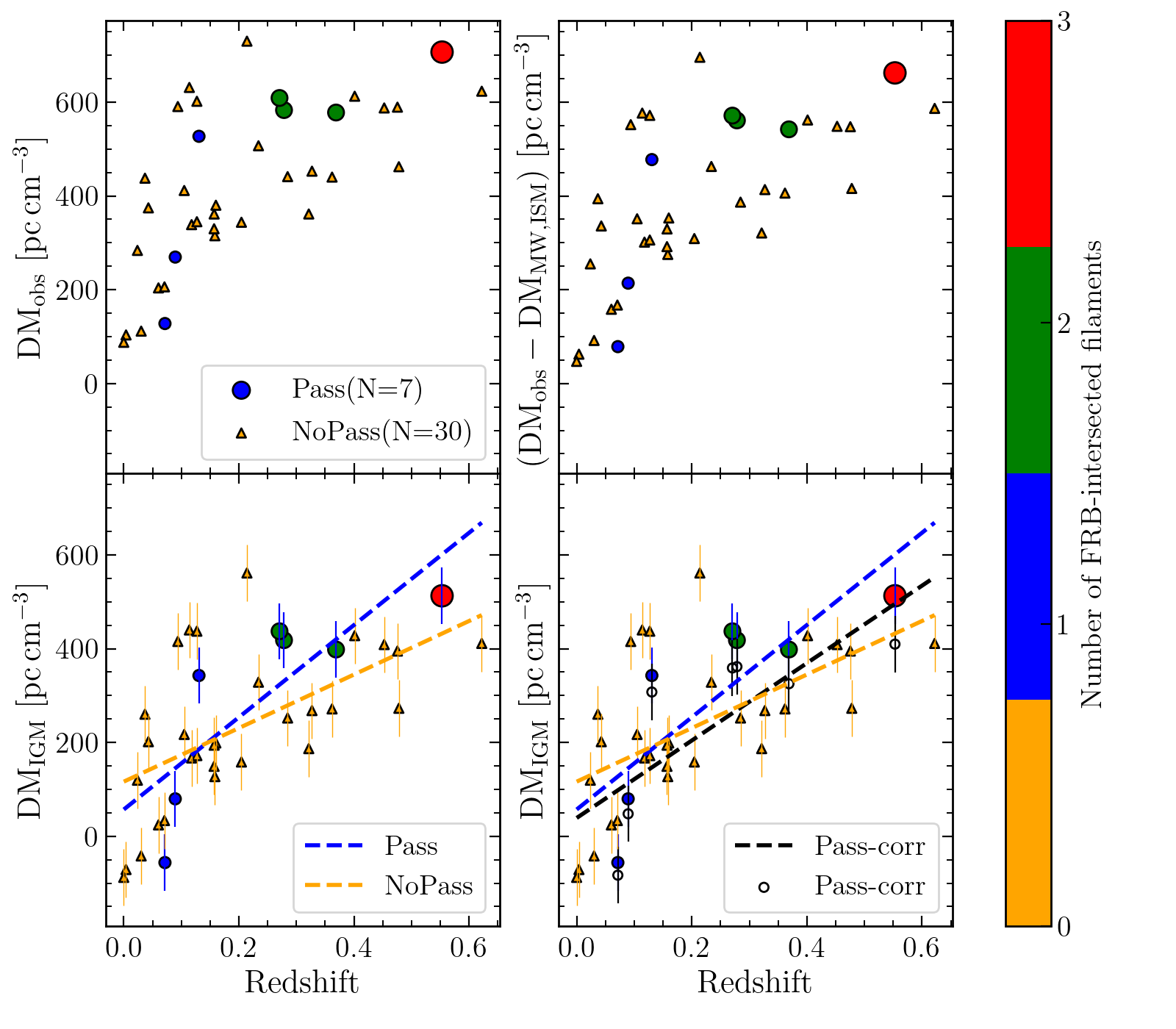}
    \caption{Top: Distributions of $\mathrm{DM_{obs}}$ (left) and $\mathrm{DM_{obs}} - \mathrm{DM_{MW,ISM}}$ (right) and redshifts for 37 well-localized FRBs in the DESI area, divided into ‘Pass’ (circles) and ‘NoPass’ (yellow triangles) groups. Circle color and size indicate the number of intersected filaments. Bottom Left: $\mathrm{DM_{IGM}}$ vs. redshift for the same FRBs. Dashed lines show best-fit linear trends for each group; error bars represent uncertainties of $\sim60\,\mathrm{pc\,cm^{-3}}$. Bottom Right: Same as left, but with filament contributions subtracted: $\mathrm{DM_{IGM,Pass\text{-}corr}} = \mathrm{DM_{IGM,Pass}} - \mathrm{DM_{fila}}$ (open circles), using central gas overdensity in filaments $\delta_0 = 20.78$. The black dashed line shows the corresponding linear fit.}
    \label{fig:M9 S5 Phost1 res}
\end{centering}
\end{figure*}

The observed DM of FRBs is the sum of multiple components:
\begin{equation}
\begin{split}
    \mathrm{DM_{obs}} &= \mathrm{DM_{MW,ISM}} + \mathrm{DM_{MW,halo}} + \mathrm{DM_{IGM}}   \\    
   &+ \frac{\mathrm{DM_{host}}}{1+z_h} + \frac{\mathrm{DM_{ForeH}}}{1+z_{fh}}  ,
    \label{eqn:DMobs sum}
\end{split}
\end{equation}
where the subscripts `MW,ISM', `MW,halo', `IGM', `Host', `ForeH' refer to the Milky Way’s interstellar medium, Milky Way halo, intergalactic medium, the host galaxy and its halo, and foreground galaxy halos, respectively. The redshifts of the host and foreground halos are denoted by $z_h$ and $z_{fh}$. The $(1+z)^{-1}$ factors account for the scaling relations between $\mathrm{DM}$ and frequency $\nu$, along with time dilation and redshift resulting from cosmic expansion \citep{2003ApJ...598L..79I}. The observed total DM and redshift distribution for the 37 well-localized FRBs are shown in the top-left panel of Figure~\ref{fig:M9 S5 Phost1 res}.

$\mathrm{DM_{IGM}}$ includes contributions from diffuse gas outside halos—voids, walls, and filaments—with the filamentary part denoted as $\mathrm{DM_{fila}}$. FRBs in the `Pass' group, which intersect filaments, are expected to have systematically higher $\mathrm{DM_{IGM}}$ than those in the `NoPass' group, especially at higher redshifts where more filaments are crossed. We estimate $\mathrm{DM_{fila}}$ from the offset between the $\mathrm{DM_{IGM}}$–$z$ relations of the two groups. $\mathrm{DM_{IGM}}$ is obtained by subtracting contributions from the Milky Way (ISM and halo), the host galaxy (and its halo), and foreground halos from the total observed DM.

We estimate $\mathrm{DM_{MW,ISM}}$ using the NE2001 model \citep{2002astro.ph..7156C}, though the YMW16 model \citep{2017ApJ...835...29Y} is an alternative. The top right panel of Figure~\ref{fig:M9 S5 Phost1 res} shows the DMs of the 37 localized FRBs after subtracting $\mathrm{DM_{MW,ISM}}$. $\mathrm{DM_{MW,halo}}$ is assumed to be a constant $\sim38\,\mathrm{pc\,cm^{-3}}$, based on \citet{2020ApJ...888..105Y} with a $68\%$ confidence range of $32$–$45\,\mathrm{pc\,cm^{-3}}$. As this value is applied uniformly, it does not affect the relative $\mathrm{DM_{IGM}}$–$z$ differences between the Pass and NoPass groups.

For the host galaxy contribution, $\mathrm{DM_{host}}$, we adopt a log-normal distribution with median $e^{\mu}_{\mathrm{host}}=130\, \mathrm{pc\,cm^{-3}}$ and shape parameter $\sigma_{\mathrm{host}}=0.56$, following \citet{2025NatAs...9.1226C}, which based on a sample of 69 localized FRBs. Following \cite{2020ApJ...900..170Z}, we assume redshift evolution as $\mathrm{DM_{host}} = 130(1 + z_h)\,\mathrm{pc\,cm^{-3}}$. For comparison, \cite{2023MNRAS.518..539M} suggested a weaker evolution, $\mathrm{DM_{host}} \propto (1 + z_h)^{0.6}$, assuming FRBs trace the cosmic star formation rate. The uncertainty, $\sigma_{\mathrm{DM_{host}}}$, is estimated from the median–16th percentile difference, yielding $\sim60\,\mathrm{pc\,cm^{-3}}$, which dominates the $\mathrm{DM_{IGM}}$ error at $z < 1$. We propagate this and adopt $\sigma_{\mathrm{DM_{IGM}}} \approx 60\,\mathrm{pc\,cm^{-3}}$. The impact of a larger $\sigma_{\mathrm{DM_{host}}}$ is explored in the fourth paragraph of Appendix ~\ref{sec:discussion_ss}. A larger $\sigma_{\mathrm{DM_{host}}}$, which corresponds to a larger $\sigma_{\mathrm{DM_{IGM}}}$, tends to alleviate the tensions in the $\mathrm{DM_{IGM}}$-$z$ relation between the `Pass' and `NoPass' groups, as shown in Table \ref{tab:models}. 

To estimate the DM from intervening foreground halos, $\mathrm{DM_{ForeH}}$, we identify candidate halos from the DESI group catalog \citep{2021ApJ...909..143Y}. A halo contributes if the FRB's L.O.S. intersects its $r_{180}$ region and no filament intersects the same L.O.S. within the halo’s redshift bin or its two adjacent bins—this avoids double-counting due to photometric redshift uncertainties and group–filament associations. For qualifying halos, we compute $\mathrm{DM_{ForeH}}$ following \cite{2025MNRAS.538.2785H}. We assume halos with $M_{\mathrm{halo}} < 10^{13.5}\,\mathrm{M_\odot}$ follow a modified NFW (mNFW) gas profile \citep{2019MNRAS.485..648P}:

\begin{equation}
% \begin{split}
    \rho_{b}(r) = f_{\mathrm{gas}}  \frac{\rho_0(M_{\mathrm{halo}})}{y^{1-\alpha}(y_0+y)^{2+\alpha}}
    \label{eqn:mNFW}
% \end{split}
\end{equation}
where $\rho_0(M_{\mathrm{halo}})$ denotes the central gas density as a function of halo mass; $y\equiv c(r/R_{\mathrm{vir}})$ with $c$ being the concentration parameter; Following \citet{2025MNRAS.538.2785H}, we fix $y_0$ and $\alpha$ to 2. The hot gas fraction, $f_{\mathrm{gas}}$, is calibrated using $R_{200}$-based scaling relations from \cite{2024arXiv241116555P}:
\begin{equation}
     f_{\mathrm{gas}} = 2.09\times10^{-6} (M_{200}/\mathrm{M_\odot}),
    \label{eqn:fgas}
\end{equation}
where $M_{200}$ is mass within $R_{200}$. For halos with $M_{\mathrm{halo}} > 10^{13.5}\,\mathrm{M_\odot}$, we adopt the ICM model from \citet{2006ApJ...640..691V}, updating the gas fraction using the same equation above. In both the mNFW (group-scale) and ICM (cluster-scale) models, the gas profile is truncated at $r_{\mathrm{max}} = R_{\mathrm{vir}}$. $\mathrm{DM_{ForeH}}$ is computed by integrating the gas density along the FRB sightline using the \texttt{Ne\_Rperp} routine from \cite{j_xavier_prochaska_2025_14804392}.

After subtracting contributions from the Milky Way, host galaxy and CGM, and foreground halos, we obtain $\mathrm{DM_{IGM}}$ and its uncertainty $\sigma_{\mathrm{DM_{IGM}}}$ for FRBs in both `Pass' and `NoPass' groups. We then estimate the baryonic mass in filaments by analyzing the difference in their $\mathrm{DM_{IGM}}$–$z$ relations. The procedure is detailed below.

We first fit a linear function, $f_{\mathrm{NoPass}}(z)$, to the $\mathrm{DM_{IGM}}$–$z$ relation for the `NoPass' subgroup, following \citet{2022ApJ...931...88C} and \citet{2025ApJS..277...43M}, who find this relation can be approximated as linear. For FRBs in `Pass' group, we assume a constant filament baryon central overdensity $\delta_0$ (10–100) and estimate $\mathrm{DM_{fila}}$ using the method in Section~\ref{subsec:DMfila}. The DM caused by the diffuse IGM in voids and walls, i.e., excluding filament contributions, is then $\mathrm{DM_{IGM,Pass\text{-}corr}} = \mathrm{DM_{IGM,Pass}} - \mathrm{DM_{fila}}$.

We use the Python package \texttt{emcee} \citep{emcee} to run Markov Chain Monte Carlo (MCMC) simulations and infer the optimal filament central overdensity $\delta_0$ by minimizing the difference between the corrected $\mathrm{DM_{IGM,Pass\text{-}corr}}$–$z$ relation and the linear fit $f_{\mathrm{NoPass}}(z)$. The adopted log-likelihood function is:
\begin{equation}
\begin{split}
    \log \mathcal{L} &=-\frac{1}{2} \sum^{N_{\mathrm{FRB,Pass}}}_{i}  (\frac{(\mathrm{DM_{IGM,Pass\text{-}corr,\mathit{i}}}-f_{\mathrm{NoPass}}(z_i))^2}{\sigma^2_i} \\
     &\quad + \log{(2\pi \sigma^2_i)} \, ),
\label{eqn:likehood}
\end{split}
\end{equation}
where $\sigma^2_i = \sigma^2_{\mathrm{Pass},i} + \sigma^2_{\mathrm{NoPass},i}$, with $\sigma^2_{\mathrm{Pass},i}$ representing the uncertainty in $\mathrm{DM_{IGM}}$ for the $i$-th FRB in the `Pass' group at redshift $z_i$, and $\sigma^2_{\mathrm{NoPass},i}$ denoting the fitting error of the linear function $f_{\mathrm{NoPass}}$  at $z_i$.

With the optimal central overdensity $\delta_0$ determined, the total gas mass in a filament is calculated as:
\begin{equation}
    \mathrm{M_{gas,fila,i}} = \int_{0}^{\mathrm{L_{fila}}} dl \int_{0}^{\mathrm{R_{fila}}} \, 2\pi r \, \rho_{\mathrm{gas}}(r) dr ,
    \label{eqn:Mgas}
\end{equation}
where $\mathrm{L_{fila}}$ is the filament length, $\mathrm{R_{fila}}$ its radius, and $\rho_{\mathrm{gas}}(r)$ the radial gas density profile.

The gas mass density from filaments in the $j$-th redshift bin is calculated by summing the gas masses of all identified filaments in that bin and dividing by the corresponding comoving volume:
\begin{equation}
    \rho_{\mathrm{gas,fila},j} = \frac{\sum_{i=1}^{\mathrm{N_{fila},j}} \mathrm{M_{gas,fila,i}}}{V_c \cdot f_{\mathrm{DESI}}},
    \label{eqn:gas density filament}
\end{equation}
where $\mathrm{N_{fila},j}$ is the number of filaments in the bin, $V_c$ is the comoving volume between redshifts $j \cdot 0.05$ and $(j+1) \cdot 0.05$, and $f_{\mathrm{DESI}} \approx 0.44$ is the DESI sky area fraction \citep{2021ApJ...909..143Y}.

For each redshift bin, the baryon fraction in filaments is $\Omega_{b,\mathrm{fila},j} = \rho_{\mathrm{gas,fila},j} / \rho_{\mathrm{crit},j}$, where $\rho_{\mathrm{crit},j}$ is the critical density at the bin’s median redshift. The average value for $z < 1$ is denoted as $\bar{\Omega}_{b,\mathrm{fila}}$.

\section{Results} \label{sec:res}
\subsection{$\mathrm{DM_{IGM}}$--$z$ relationships of two FRB groups}

Among 37 securely localized FRBs in the DESI area, 7 intersect filaments identified by DisPerSE using galaxies with $M_* > 10^9,\mathrm{M_{\odot}}$ and \texttt{nsig=5}, while 30 do not (see Figure \ref{fig:M9 S5 Phost1 res}). We quantify the difference by fitting separate linear regressions to the $\mathrm{DM_{IGM}}$–$z$ relations for the `Pass' and `NoPass' groups. The best-fit relations are:

\begin{equation}
    \mathrm{DM_{IGM,Pass}} = a_1 z+b_1= (982\pm136)\cdot z+(58\pm40),
    \label{eqn:dmPass}
\end{equation}
\begin{equation}
    \mathrm{DM_{IGM,NoPass}} = a_2 z+b_2 = (573\pm65)\cdot z+(116\pm16).
    \label{eqn:dmNoPass}
\end{equation}

The parameter covariance matrices are
$\mathrm{cov}_1=\left( \begin{smallmatrix} 18461 & -4656 \\ -4656 & 1644 \end{smallmatrix} \right)$ and $\mathrm{cov}_2=\left( \begin{smallmatrix} 4229 & -838 \\ -838 & 276 \end{smallmatrix} \right)$. The difference between the fit parameters is given by $\mathbf{D} = (a_1 - a_2,, b_1 - b_2)$, and the chi-square statistic is $\chi^2=\mathbf{D} \mathrm{(cov_1+cov_2)}^{-1}\mathbf{D}^{T} = 10.29$, yielding a p-value of 0.006, or $\sim 2.8\,\sigma$, assuming a chi-square distribution with 2 degrees of freedom. This result indicate a statistically significant $2.8\,\sigma$ difference in the $\mathrm{DM_{IGM}}$–$z$ relation between the `Pass' and `NoPass' groups.

\citet{2022ApJ...931...88C} found $\mathrm{DM_{IGM}} \approx 826z\,\mathrm{pc\,cm^{-3}}$ assuming a constant IGM fraction $f_{\mathrm{IGM}} = 0.85$, while \citet{2025ApJS..277...43M} reported $\mathrm{DM_{IGM}} = 861z - 25\,\mathrm{pc\,cm^{-3}}$ using a redshift-dependent $f_{\mathrm{IGM}}(z)$ from TNG100. These values fall between our `Pass' and `NoPass' results—higher than the latter and lower than the former—as expected, since they reflect the average IGM contribution, including diffuse gas in voids, walls, and filaments.

As shown in the bottom left panel of Figure~\ref{fig:M9 S5 Phost1 res}, the $\mathrm{DM_{IGM}}$–$z$ relations for the `Pass’ and `NoPass' groups diverge at higher redshifts, as distant FRBs are more likely to intersect filaments, increasing their $\mathrm{DM}$. This difference becomes more significant with larger samples—reaching $4.9\,\sigma$ with all 46 FRBs in the DESI area, and up to $5.9\,\sigma$ when considering only those at $z > 0.1$. Using a lower persistence threshold (\texttt{nsig=3}) reduces the significance to $2.4\,\sigma$ (37 FRBs) and $3.5\,\sigma$ (46 FRBs). Details on the factors affecting this significance, including number of localized FRBs, the redshift range of selected samples, the assumed $\sigma_{\mathrm{DM_{host}}}$, and filaments classification procedure are provided in Appendix~\ref{sec:discussion_ss}.

\subsection{Baryon mass in filaments}
Using the method in Section~\ref{subsec:estimate baryon}, we perform MCMC sampling with 10 walkers, 500 burn-in steps, and 5000 total steps to estimate the optimal central baryon overdensity in filaments. Convergence, typically achieved when the chain length exceeds $50\,\tau_f$ \citep{emcee}, was reached after $\sim 4000$ steps in our case.

The MCMC analysis yields an optimal central overdensity of $\delta_0 = 21^{+13}_{-12}$, assuming an isothermal single-$\beta$ model with $\beta = 2/3$. The corrected $\rm{DM_{IGM}}$, $\mathrm{DM_{IGM,Pass\text{-}corr}} = \mathrm{DM_{IGM,Pass}} - \mathrm{DM_{fila}}$, are shown in the bottom right panel of Figure~\ref{fig:M9 S5 Phost1 res}, with $\mathrm{DM_{fila}}$ values listed in Table~\ref{tab:FRBs}. On average, each intersected filament contributes about $34\,\mathrm{pc\,cm^{-3}}$.

 \begin{figure}
    \centering
    \includegraphics[width=1.\columnwidth]{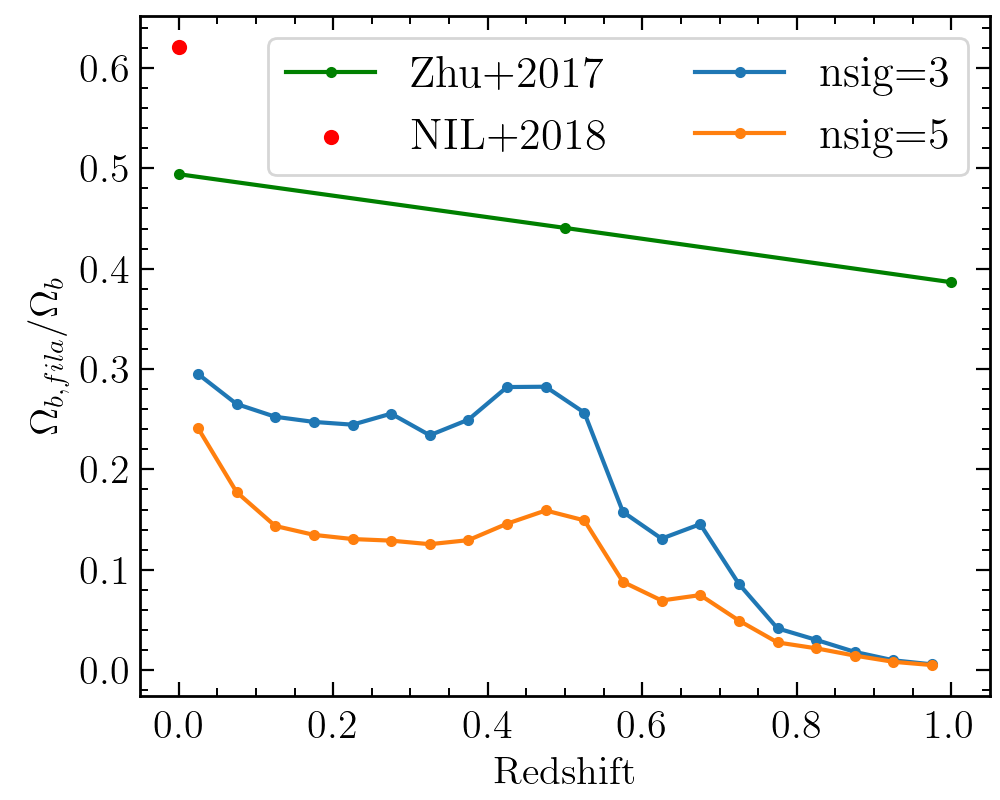}
    \caption{Estimated baryon mass fraction in filaments across redshift bins. Blue and orange lines show results from this work using 37 securelylocalized FRBs within DESI area and DisPerSE-identified filaments (galaxy catalog from DESI imaging surveys) with persistence thresholds \texttt{nsig=3} and \texttt{nsig=5}, respectively. The green line and red point represent estimates from simulations by \cite{2017ApJ...838...21Z} and \cite{2018MNRAS.473.1195L}.} 
    \label{fig:omegaB fila}
\end{figure}

This best-fit central overdensity is broadly consistent with predictions from hydrodynamical simulations (e.g., \citealt{2017ApJ...838...21Z, 2019MNRAS.486.3766M, 2021A&A...646A.156T, 2021ApJ...920....2Z,2022A&A...661A.115G,2025ApJ...989..187Y}) and aligns well with several observational estimates. It agrees with $\delta_0 \sim 19$ from tSZ measurements of filaments identified in SDSS DR12 \citep{2020A&A...637A..41T} and $\delta_0 \sim 21$ from X-ray emission measurements of filaments using SRG/eROSITA \citep{2022A&A...667A.161T}, but is higher than the $\delta_0 \sim 3.2$–$5.5$ values reported by \cite{2019MNRAS.483..223T} and \cite{2019A&A...624A..48D} based on tSZ measurements of filaments between galaxy group pairs.

Using Equations~\ref{eqn:Mgas} and~\ref{eqn:gas density filament}, we derive the baryon mass fraction in filaments for each redshift bin, shown in Figure~\ref{fig:omegaB fila}. With \texttt{nsig=5} in filament identification, the fraction declines from $0.25\,\Omega_b$ at $z=0.02$ to $0.15\,\Omega_b$ at $z=0.5$, and $0.03\,\Omega_b$ at $z=0.8$. Lowering the threshold to \texttt{nsig=3} raises these to $0.30$, $0.30$, and $0.04$, respectively. The bump in $\Omega_{b,\mathrm{fila}}$ between $z = 0.4$ and $0.55$ is due to a higher filament count in that range (Figure~\ref{fig:exampleFila}). Averaged over $z < 1$, filaments contribute $\Omega_{b,\mathrm{fila}} = 0.10\,\Omega_b$ for \texttt{nsig=5} and $0.17\,\Omega_b$ for \texttt{nsig=3}.

Our measurement agrees with the tSZ-based estimate from \cite{2020A&A...637A..41T}, who found $\Omega_{b,\mathrm{fila}} = 0.08\,\Omega_b$ for SDSS filaments at $0.2 < z < 0.6$. More recently, Li et al. (submitted) used Planck Compton-$y$ and CMB lensing maps to estimate $\Omega_{b,\mathrm{fila}} = 0.12^{+0.020}_{-0.021},\Omega_b$ at $0.2 < z < 0.6$ for long filaments (30–100 $h^{-1}\,\mathrm{Mpc}$) identified from SDSS DR12, and $0.227^{+0.036}_{-0.035}\,\Omega_b$ for filaments of all lengths.

Even with \texttt{nsig=3}, the derived filament baryon fractions at $z < 1$ are still below the $30\%$–$60\%$ predicted by simulations (e.g., \citealt{2017ApJ...838...21Z, 2018MNRAS.473.1195L}). This discrepancy likely stems from the incompleteness of our filament sample, limited by the galaxy density in the DESI imaging catalog—especially at $z > 0.5$, where many filaments may be missed.

\section{Summary}
\label{sec:summary}
We classify fast radio bursts (FRBs) into two groups based on whether their lines of sight intersect cosmic filaments (`Pass') or not (`NoPass'), with filaments identified using the DisPerSE algorithm applied to galaxies with $M_* > 10^9,\mathrm{M_{\odot}}$ from DESI imaging surveys.
We find tentative ($\sim 3\,\sigma$) evidence for a divergence in the $\mathrm{DM_{IGM}}$–$z$ relation between the two groups. Assuming this divergence arises from ionized baryons in filaments and modeling their gas with an isothermal $\beta$-profile ($\beta=2/3$), MCMC analysis yields a central baryon overdensity of $\delta = 21^{+13}_{-12}$.

The inferred central overdensity broadly aligns with both hydrodynamical simulations and observational estimates from tSZ and X-ray data. Using this value, we estimate that filaments host $\Omega_{b,\mathrm{fila}} = 0.10$–$0.17\,\Omega_b$ at $z < 1$. The fraction declines from $0.25$–$0.30\,\Omega_b$ at $z = 0.02$ to $0.15$–$0.30\,\Omega_b$ at $z = 0.5$, and $0.03$–$0.04\,\Omega_b$ at $z = 0.8$. These values likely represent lower limits due to catalog incompleteness, especially at $z>0.5$.

This study offers an independent method to trace baryons in cosmic filaments and refine their density estimates, contributing to resolving the missing baryon problem. Our findings underscore the need for larger samples of localized FRBs and galaxies, along with complementary probes like X-ray emission and the Sunyaev-Zel’dovich effect, for further validation and improvement. 

We performed a simplified random modeling analysis and found that approximately 90–100 high-confidence localized FRBs are needed to reach a $5 \sigma$ tension in the $\mathrm{DM_{IGM}}$–$z$ relation between the `Pass' and `NoPass' groups, if $\rm{\sigma_{DM_{IGM}}=100\,pc\,cm^{-3}}$, which is more conservative than the $\rm{60\,pc\,cm^{-3}}$ adopted in our default model. We generate mock FRBs with redshifts drawn from the distribution of all localized FRBs in the DESI region. A Pass-to-NoPass ratio of 1:4 was adopted for mock events, which is similar to the observed ratio of 7:30 among the 37 securely localized FRBs.
Their $\rm{DM_{IGM}}$ values were assigned by sampling from log-normal distributions with variance $\rm{\sigma^2_{DM_{IGM}}}$ and mean values determined by the corresponding fitted $\mathrm{DM_{IGM}}$–$z$ relations derived from these 37 FRBs under $\rm{\sigma_{DM_{IGM}}=100\,pc\,cm^{-3}}$.  

% \acknowledgments
\begin{acknowledgments}
We thank the anonymous referee for his/her valuable suggestions, which have helped improve the manuscript. We thank Zhi-Qi Huang for his helpful discussion. This work is supported by the National Natural Science Foundation of China (NFSC) through grant 11733010, 12173102, and 12203107, and by the China Manned Space Program through its Space Application System.. This work has used the HPC facility of the School of Physics and Astronomy, Sun Yat-Sen University.\\

\end{acknowledgments}

\begin{contribution}
WSZ conceived the initial research concept and was responsible for writing and submitting the manuscript. JFM conducted the formal analysis and validation and also contributed to the initial draft. QRY, YZ, and LLF contributed to both the analysis and the development of the research concept.
\end{contribution}


%%\vspace{30mm}
%facilities{HST(STIS), Swift(XRT and UVOT), AAVSO, CTIO:1.3m,


\software{DisPerSE \citep{2011MNRAS.414..350S, 2011MNRAS.414..384S}
%          Cloudy \citep{2013RMxAA..49..137F}, 
%          SExtractor \citep{1996A&AS..117..393B}
         }
% 2011MNRAS.414..350S, 2011MNRAS.414..384S
\vspace{15mm}



\appendix 
\section{filament classification}
\label{sec:appendix_class}

The details for how we identify filaments from the galaxy catalog in DESI imaging survey using DisPerSE algorithm are as follows.
We first compute the 2D density field of the galaxy distribution using the \texttt{delaunay\_2D} function in DisPerSE, which implements the Delaunay Tessellation Field Estimator (DTFE; \citealt{2000A&A...363L..29S, 2009LNP...665..291V}). We select galaxies with stellar masses $M_*>10^{9} \mathrm{M_\odot}$, a mass cut consistent with a number of previous studies (e.g.  \citealt{2024A&A...684A..63G,2024MNRAS.534.1682O,2025ApJ...989..187Y}). To deal with boundary conditions in the observational sample area, \cite{2022MNRAS.517.1678C} generated a random uniform distribution of artificial galaxies beyond the boundary. Similarly, we utilize the \texttt{-btype smooth} parameter in the \texttt{delaunay\_2D} function, consistent with \cite{2020A&A...642A..19M}. This approach adds supplementary particles outside the boundary by interpolating the estimated density within boundary. 

Next, we smooth the density field one time using the \texttt{netconv} function, which averages the density values with those of their direct neighbors in the network. Subsequently, we extract filaments from the smoothed galaxy density field using the \texttt{mse} function, which implements discrete Morse theory. This involves computing the gradients of the density field and identifying critical points (minima, saddles, and maxima) where the gradient vanishes.
A filament is defined by the \texttt{skelconv} function as two field lines originating from a saddle point and connecting two maxima. The resulting filament skeleton remains unsmoothed in this work.

Spurious filaments, likely arising from Poisson noise in the discrete galaxy distribution, are eliminated using persistence theory within DisPerSE. 
Following \cite{2020A&A...642A..19M}, we set the \texttt{nsig} parameter in the \texttt{mse} function to 3 or 5. This means structures with a probability less than 3$\sigma$ or 5$\sigma$ of appearing in a random field are removed. Consequently, filaments extracted with a higher persistence threshold are more robust than those extracted with a lower threshold, but may lose real weak filaments. With due caution, we adopt results with a 5$\sigma$ (\texttt{nsig}=5) persistence threshold as our default sample of filaments, which will compared with results with \texttt{nsig}=3 to inspect the impact. 

As an example, the middle panel in Figure \ref{fig:exampleFila} show filaments identified within redshift range $0-0.05$ using a \texttt{nsig}=3 and \texttt{nsig}=5, respectively, alongside the galaxy distribution. The bottom panel in Figure \ref{fig:exampleFila} shows the numbers of galaxies and filaments we identified in each redshift bin.

\section{Procedures of identifying intersected filaments}
\label{sec:appendix_intersect}

After finding candidate intersected filaments, we estimate their physical radius, $\mathrm{R_{fila}}$, based on the correlation between the $\mathrm{R_{fila}}$ and the stellar mass per unit length contained within the filament found by \cite{2025ApJ...989..187Y}. Based on the Illustris-TNG simulation, \cite{2025ApJ...989..187Y} find that $\mathrm{R_{fila}}$ correlates with the stellar mass per unit length contained within the filament, denoted as $\mathrm{M_{\ast}/L_{fila}}$. In this context, $\mathrm{R_{fila}}$ is defined as the radial distance from the filament spine at which the gas density drops to the mean cosmic baryon density. The relationship between $\mathrm{R_{fila}}$ and $\mathrm{M_{\ast}/L_{fila}}$ is provided by fitting results at redshifts 0, 0.5, 1.0, and 2.0 as follows:

\begin{equation}
\begin{split}
     \mathrm{log_{10}(R_{fila})} &= \mathrm{\frac{log_{10}(M_*/L_{fila})+1.58}{2.535}}, z=0 \\
     \mathrm{log_{10}(R_{fila})} &= \mathrm{\frac{log_{10}(M_*/L_{fila})+0.285}{2.233}}, z=0.5 \\
     \mathrm{log_{10}(R_{fila})} &= \mathrm{\frac{log_{10}(M_*/L_{fila})+0.26}{2.364}}, z=1 \\
     \mathrm{log_{10}(R_{fila})} &= \mathrm{\frac{log_{10}(M_*/L_{fila})-3.209}{1.412}}, z=2 \\
    \label{eqn:filament radius vs M}
\end{split}
\end{equation}
where $\mathrm{L_{fila}}$ is the filament length in comoving kiloparsecs (ckpc), while $\mathrm{R_{fila}}$ and $M_{\ast}$ are given in ckpc and solar masses ($\mathrm{M}_{\odot}$), respectively. To estimate $\mathrm{R_{fila}}$ at arbitrary redshifts,we apply linear interpolation between the fitting relations at $z=0$, 0.5, 1.0, and 2.0.

Using the filament skeleton determined by DisPerSE as the central axis, we calculate the total stellar mass within a cylindrical volume of initial radius 2 Mpc from the galaxy catalog. From this, we obtain an initial estimate of the filament radius via Equation~\ref{eqn:filament radius vs M}. This radius is then iteratively refined by repeating the procedure: updating the cylindrical radius and recalculating the enclosed stellar mass. The iteration proceeds until the relative change in radius between steps falls below $5\%$, or a maximum of 10 iterations is reached. The final value is adopted as $\mathrm{R_{fila}}$, with an upper limit of 5 Mpc imposed to avoid overestimation. It is worth noting that limited galaxy sampling in the DESI imaging survey at high redshift may lead to underestimated filament radii.

With the improved estimation of filament widths, we then verify whether FRB line of sight (L.O.S.) intersects the candidate filaments in three-dimensional space, and subsequently compute the corresponding $\mathrm{DM_{fila}}$. This process involves converting the FRB and filament positions from sky coordinates to Cartesian coordinates using the following transformation:
\begin{equation}
\begin{split}
     \mathrm{X} &= r_{\mathrm{com}} \cdot \cos(\mathrm{RA}) \cdot \cos(\mathrm{DEC}), \\
     \mathrm{Y} &= r_{\mathrm{com}} \cdot \sin(\mathrm{RA}) \cdot \cos(\mathrm{DEC}), \\
     \mathrm{Z} &= r_{\mathrm{com}} \cdot \sin(\mathrm{DEC}).
    \label{eqn:sky to Cart}
\end{split}
\end{equation}

An FRB is considered to intersect a filament if its line of sight passes through the filament’s cylindrical region of radius $\mathrm{R_{fila}}$ in three-dimensional space. The intersection path is shown as the red segment in the bottom panel of Figure~\ref{fig:through fila 2d}. 

\section{factors may influence the statistical significance} 
\label{sec:discussion_ss}
\begin{figure*}
    \centering
    \includegraphics[width=0.45\textwidth]{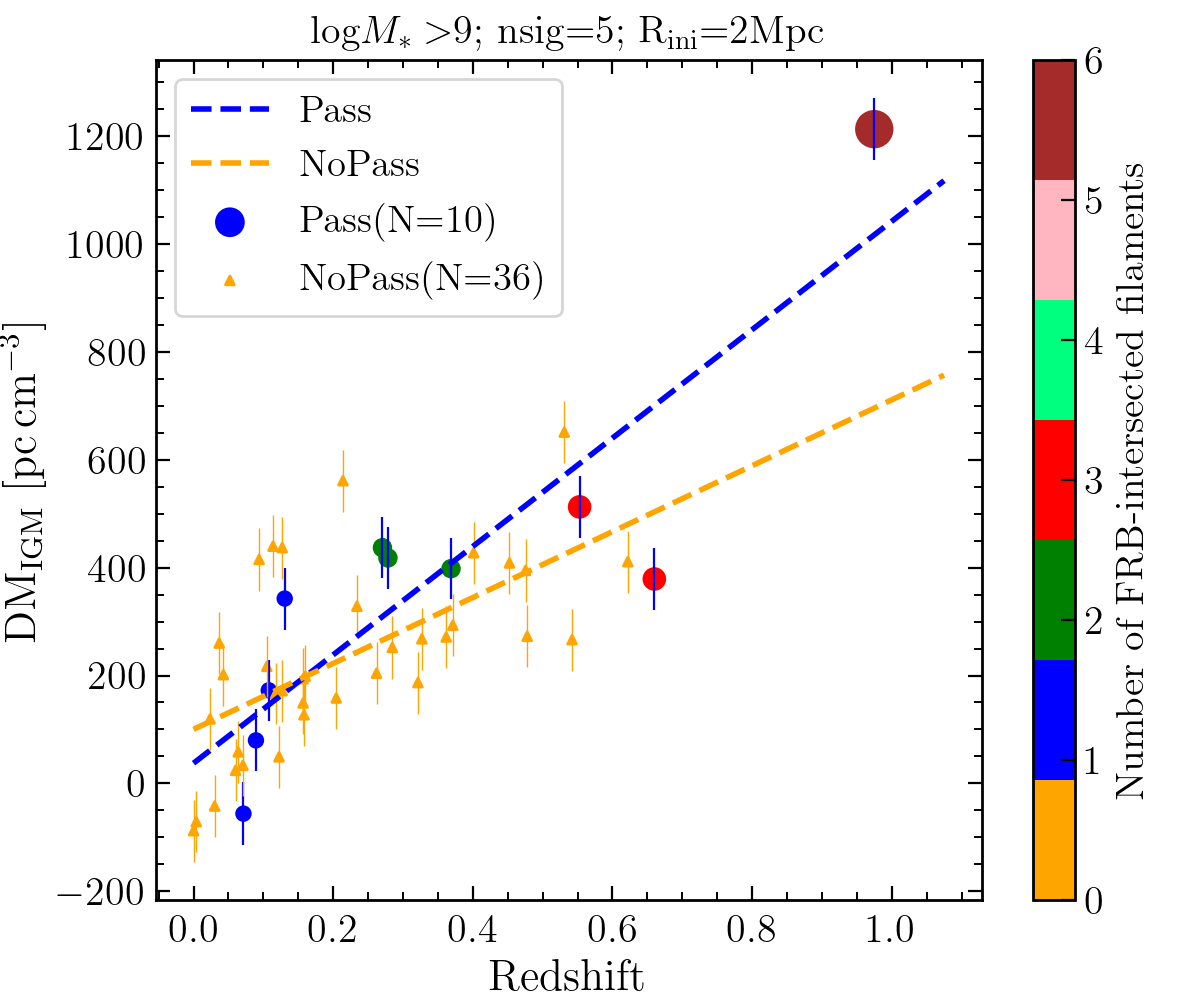}
    \includegraphics[width=0.45\textwidth]{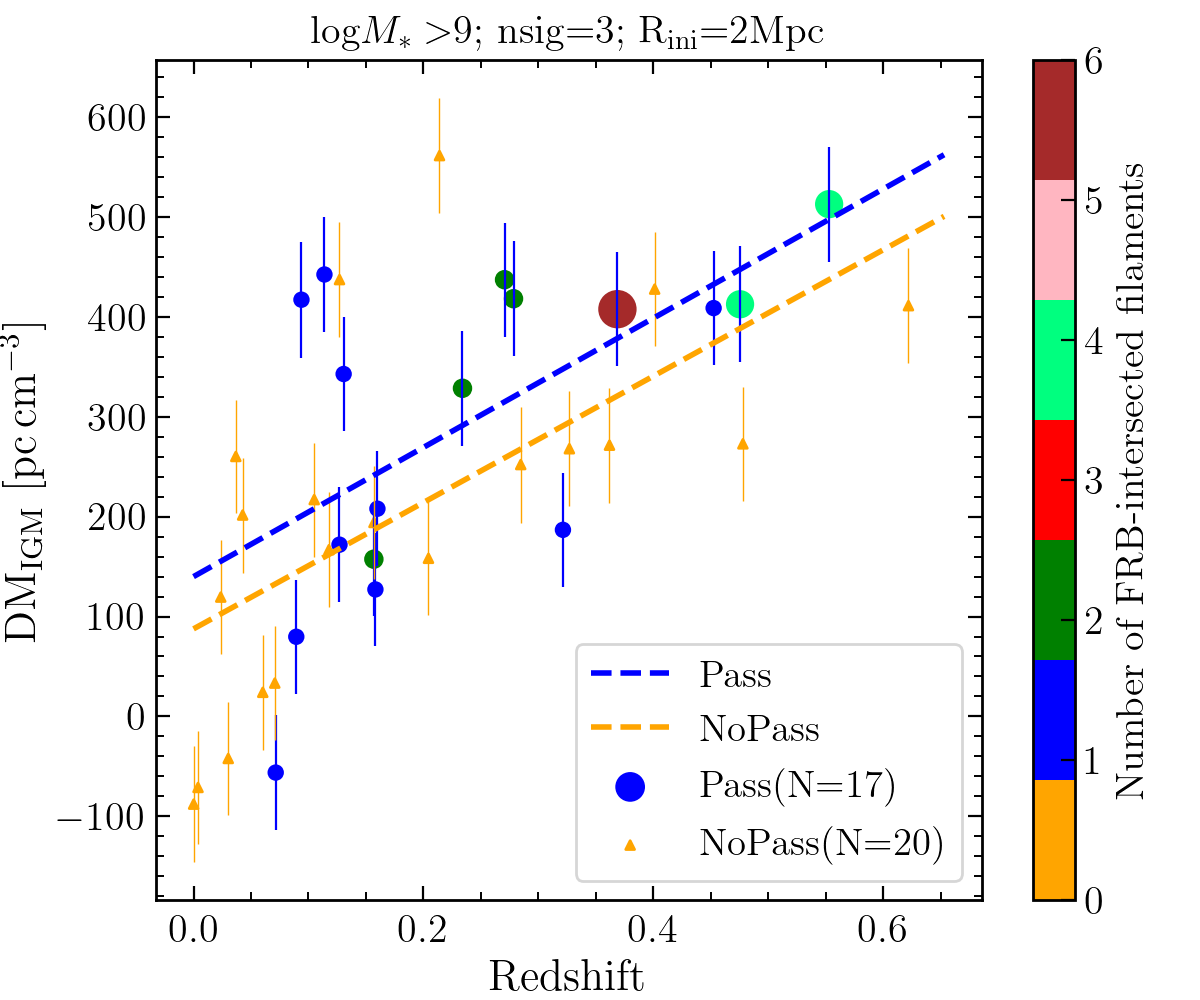}
    \caption{Similar to Figure \ref{fig:M9 S5 Phost1 res}, the left panel presents results based on all 46 localized FRBs within the DESI sky area, using filaments identified by DisPerSE with a persistence threshold of \texttt{nsig=5}. In contrast, the right panel shows results for the 37 securely localized FRBs, with filaments extracted using a lower threshold of \texttt{nsig=3}. }
    \label{fig:M9 S5 phost0 and M9 S3} 
\end{figure*}

Several factors may influence the statistical significance of our results, including the number and reliability of localized FRBs, the redshift distribution of the selected sample, the uncertainty in $\mathrm{DM_{IGM}}$ ($\sigma_{\mathrm{DM_{IGM}}}$), the method used to classify filaments, and the assumed value of the initial filament radius $\mathrm{R_{ini}}$. We examine these aspects below.

Increasing the number of well-localized FRBs is essential to enhance the robustness of our conclusions. As illustrated in the left column of Figure \ref{fig:M9 S5 phost0 and M9 S3}, when all localized FRBs within the DESI sky area ($\mathrm{N_{FRB}} = 46$) are included, the difference in the $\mathrm{DM_{IGM}}$–$z$ relations between the `Pass’ and `NoPass’ groups reaches a statistical significance of $4.9\,\sigma$.

Focusing exclusively on localized FRBs at intermediate to high redshifts can improve the statistical significance of our results, as the relative contribution of the IGM to the total DM increases with redshift compared to nearby events. To evaluate this effect, we repeat our analysis using securely localized FRBs with $z > 0.1$ ($\mathrm{N_{FRB}} = 26$) and \texttt{nsig}=5, while keeping all other procedures unchanged. This yields $3.7\,\sigma$ evidence for a divergent trend in the $\mathrm{DM_{IGM}}$–$z$ relation between the `Pass’ and `NoPass’ groups. When applying the same analysis to the full sample of localized FRBs within the DESI sky coverage at $z > 0.1$ ($\mathrm{N_{FRB}} = 34$), the statistical significance increases to $5.9\,\sigma$. Therefore, the accumulation of additional localized FRBs at $z > 0.1$ in the future will be crucial for strengthening the statistical robustness of the $\mathrm{DM_{IGM}}$–$z$ relations and further testing our findings.

Our default model adopts an uncertainty of $\sigma_{\mathrm{DM_{IGM}}} \sim 60\,\mathrm{pc\,cm^{-3}}$ \citep{2025NatAs...9.1226C}, yielding a $2.8\sigma$ level of significance for the securelylocalized FRB sample ($\mathrm{N_{FRB}}=37$). Increasing $\sigma_{\mathrm{DM_{IGM}}}$ to $\sim100\,\mathrm{pc\,cm^{-3}}$ reduces the significance to $1.3\,\sigma$. Considering an even larger uncertainty based on the redshift-evolving $\mathrm{DM_{host}}$ model from TNG simulations by \cite{2023MNRAS.518..539M}, which suggests $\mathrm{DM_{host}} \sim 170$–$220\,\mathrm{pc\,cm^{-3}}$, the significance further decreases to $0.5\,\sigma$. The relatively large uncertainty in $\mathrm{DM_{host}}$ arises from the diversity of host galaxy types among localized FRBs, including both star-forming and quiescent systems, as well as spiral and elliptical morphologies. However, the impact of this large $\sigma_{\mathrm{DM_{host}}}$—and consequently the uncertainty in $\sigma_{\mathrm{DM_{IGM}}}$—is expected to decline as the number of localized FRBs increases. For instance, applying the same analysis to all localized FRBs within the DESI sky coverage ($\mathrm{N_{FRB}}=46$) yields improved significance levels of $2.6\,\sigma$ and $1.4\,\sigma$ when adopting $\sigma_{\mathrm{DM_{IGM}}} = 100$ and $170$–$200\,\mathrm{pc\,cm^{-3}}$, respectively—demonstrating that a larger sample size can mitigate the effects of increased uncertainty in $\mathrm{DM_{host}}$ and $\mathrm{DM_{IGM}}$.

We also evaluate the impact of using filaments extracted with a lower persistence threshold of \texttt{nsig=3}, compared to the default \texttt{nsig=5}, based on the 37 securelylocalized FRB samples and assuming $\sigma_{\mathrm{DM_{IGM}}} \sim 60\,\mathrm{pc\,cm^{-3}}$. This lower threshold increases the number of FRBs whose sightlines intersect filaments—including those deemed less significant—raising the count to 17, in contrast to 7 under the \texttt{nsig=5} setting (see the right panel of Figure \ref{fig:M9 S5 phost0 and M9 S3}). Under these conditions, the statistical significance of the difference in the $\mathrm{DM_{IGM}}$–$z$ relation between the `Pass’ and `NoPass’ groups reaches $2.4\,\sigma$. Furthermore, when the full set of 46 localized FRBs within the DESI sky coverage is used, the corresponding significance increases to $3.5\,\sigma$.

The choice of input galaxy data influences the reconstructed density field and, consequently, the filament extraction. We test an alternative selection by using galaxies with stellar masses $M_* > 10^{10}\,\mathrm{M_{\odot}}$ to identify filaments and find a statistical significance of $3.2\,\sigma$ for both \texttt{nsig=5} and \texttt{nsig=3} persistence thresholds. Additionally, we examine the impact of the adopted initial guess for the filament width, $\mathrm{R_{ini}}$. When setting $\mathrm{R_{ini}} = 5$ Mpc, the significance of the difference between the `Pass' and `NoPass' groups drops to $1.4\,\sigma$ for the securelylocalized FRB sample ($\mathrm{N_{FRB}} = 37$), and to $2.9\,\sigma$ for the full localized FRB sample within the DESI sky area ($\mathrm{N_{FRB}} = 46$). A summary of all these tests is provided in Table \ref{tab:models}.

\begin{deluxetable*}{c|c|c|cc|cc}
\tablecaption{Summary of model parameters and inferred results. `Tension' refers to the statistical significance of the difference in the $\mathrm{DM_{IGM}}$–$z$ relationships between FRBs in the `Pass' and `NoPass' groups. `Overdensity' denotes the best-fit central baryonic overdensity ($\delta_0$) in filaments.}
\label{tab:models}
\tablehead{
\multicolumn{5}{c|}{model parameters} & \multicolumn{2}{c}{results}  \\
\colhead{galaxy stellar mass cut} & \colhead{persistence threshold} & \colhead{$\mathrm{R_{ini}}$[Mpc]}  & \colhead{$\sigma_{\mathrm{DM_{IGM}}}$ $[\mathrm{pc\,cm^{-3}}]$} &  \multicolumn{1}{c|}{FRB number} & \colhead{tension} & \colhead{central overdensity $\delta_0$}}  %\colhead{FRB number}

\startdata
\multirow{10}{*}{$>10^9\mathrm{M_{\odot}}$} & \multirow{8}{*}{5$\sigma$} & \multirow{6}{*}{2} & \multirow{4}{*}{60} & 37 & 2.8$\sigma$ & $21^{+13}_{-12}$  \\ 
&  &  &  & 46 & 4.9$\sigma$ & $22^{+8}_{-8}$  \\ 
&  &  &  & 26 ($z>0.1$) & 3.7$\sigma$ & $36^{+16}_{-17}$  \\ 
&  &  &  & 34 ($z>0.1$) & 5.9$\sigma$ & $30^{+9}_{-9}$  \\ 
\cline{4-7}
&  &  & \multirow{2}{*}{100} & 37 & 1.3$\sigma$ & $24^{+20}_{-15}$  \\ 
&  &  &  & 46 & 2.6$\sigma$ & $24^{+12}_{-11}$  \\ 
\cline{4-7}
&  &  & \multirow{2}{*}{170-220} & 37 & 0.6$\sigma$ & $37^{+36}_{-25}$  \\ 
&  &  &  & 46 & 1.4$\sigma$ & $31^{+23}_{-20}$  \\ 
\cline{3-7}
&  & \multirow{2}{*}{5} & \multirow{2}{*}{60} & 37 & 1.4$\sigma$ & $25^{+11}_{-11}$  \\ 
&  &  &  & 46 & 2.9$\sigma$ & $24^{+8}_{-7}$  \\ 
\cline{2-7}
& \multirow{2}{*}{3$\sigma$} & \multirow{2}{*}{2} & \multirow{2}{*}{60} & 37 & 2.4$\sigma$ & $13^{+7}_{-6.}$  \\ 
&  &  &  & 46 & 3.5$\sigma$ & $17^{+5}_{-6}$  \\ 
\cline{1-7}
\multirow{2}{*}{$>10^{10}\mathrm{M_{\odot}}$} & \multirow{1}{*}{5$\sigma$} & \multirow{1}{*}{2} & \multirow{1}{*}{60} & 37 & 3.2$\sigma$ & $49^{+19}_{-18}$  \\ 
& \multirow{1}{*}{3$\sigma$} & \multirow{1}{*}{2} & \multirow{1}{*}{60} & 37 & 3.2$\sigma$ & $30^{+8}_{-9}$  \\ 
% &  &  &  & 46 & 4.9$\sigma$ & $22.49^{+7.97}_{-7.79}$  \\ 
\enddata
\end{deluxetable*}

%% For this sample we use BibTeX plus aasjournalv7.bst to generate the

\bibliography{main}
\bibliographystyle{aasjournalv7}



\end{document}